\documentclass[11pt]{article}

\usepackage{graphicx}
\usepackage{amsmath, amsthm, amssymb}
\usepackage{subfigure}
\usepackage{comment}

\newcommand{\R}{\mathbb{R}}

\newcommand{\C}{\mathbb{C}}

\newcommand{\E}{\mathbb{E}}

\newcommand{\ket}[1]{| #1 \rangle}
\newcommand{\bra}[1]{\langle #1|}
\newcommand{\ip}[2]{\langle #1|#2 \rangle}
\newcommand{\bracket}[3]{\langle #1|#2|#3 \rangle}
\newcommand{\sm}[1]{\left( \begin{smallmatrix} #1 \end{smallmatrix} \right)}

\newcommand{\be}{\begin{equation}}
\newcommand{\ee}{\end{equation}}
\newcommand{\bea}{\begin{eqnarray}}
\newcommand{\eea}{\end{eqnarray}}
\newcommand{\bes}{\begin{equation*}}
\newcommand{\ees}{\end{equation*}}
\newcommand{\beas}{\begin{eqnarray*}}
\newcommand{\eeas}{\end{eqnarray*}}

\newtheorem{thm}{Theorem}[section]
\newtheorem*{thm*}{Theorem}

\newtheorem{lem}[thm]{Lemma}
\newtheorem*{lem*}{Lemma}

\newtheorem*{lipschitzLem*}{Lemma \ref{lipschitz}}
\newtheorem*{lipschitzCubeLem*}{Lemma \ref{lipschitzCube}}
\newtheorem*{pgmNearlyOptimalThm*}{Theorem \ref{pgmNearlyOptimal}}

\begin{document}

\title{On the distinguishability of random quantum states}

\date{\today}

\author{Ashley Montanaro\footnote{montanar@cs.bris.ac.uk}\\ \\ {\small Department of Computer Science, University of Bristol,}\\{\small Woodland Road, Bristol, BS8 1UB, UK.} }

\maketitle

\begin{abstract}
We develop two analytic lower bounds on the probability of success $p$ of identifying a state picked from a known ensemble of pure states: a bound based on the pairwise inner products of the states, and a bound based on the eigenvalues of their Gram matrix. We use the latter to lower bound the asymptotic distinguishability of ensembles of $n$ random quantum states in $d$ dimensions, where $n/d$ approaches a constant. In particular, for almost all ensembles of $n$ states in $n$ dimensions, $p>0.72$. An application to distinguishing Boolean functions (the ``oracle identification problem'') in quantum computation is given.
\end{abstract}


\section{Introduction}

A fundamental property of quantum mechanics is that non-orthogonal pure quantum states may not be distinguished perfectly. This leads to the following {\em quantum detection problem}: given an unknown quantum state $\ket{\psi_?}$, picked from a known set $\mathcal{E}$ with known a priori probabilities, find the ``optimal'' measurement $M^{opt}$ to determine $\ket{\psi_?}$. Several different criteria for optimality may been considered \cite{helstrom,davies,eldar2}; here we only concern ourselves with optimising the probability of success $P^{opt}$, and in particular the related {\em state distinguishability problem} of finding $P^{opt}$ without necessarily finding $M^{opt}$. Efficient optimisation techniques can be used to estimate $P^{opt}$ numerically \cite{eldar}; however, the problem of finding an analytic expression for $P^{opt}$ seems intractable. We are therefore led to attempting to produce bounds on $P^{opt}$.

This note derives two lower bounds on $P^{opt}$; one based on the pairwise distinguishability of the states in $\mathcal{E}$, and one based on the eigenvalues of their Gram matrix. We use the latter, and a powerful result from random matrix theory (the Mar\v{c}enko-Pastur law \cite{marchenko}), to bound the probability of distinguishing a set of random quantum states, for a quite general notion of randomness. This has an application to quantum computation in the so-called oracle identification problem introduced by Ambainis et al \cite{ambainis}, where we are given an $n$-bit Boolean function $f$ picked from a known set of $N$ functions, and must identify $f$ with the minimum number of queries to $f$. We show that, for all but an exponentially small fraction of sets with $N=2^n$, a quantum computer can perform this task successfully in a constant number of queries (with arbitrarily high probability), whereas classical computation requires $n$ queries for all such sets.

As showing that a set of quantum states are quite distinguishable forms an essential part of proofs in many areas of quantum information theory, we hope that these results will find application elsewhere.

The organisation of the paper is as follows. Section 2 introduces notation and our main tool, the so-called ``pretty good measurement'', before moving on to give the lower bounds on $P^{opt}$. An extension of the lower bounds to mixed states is considered. Section 3 applies the bounds to a specific family of ensembles (those where all the states have constant inner product). Section 4 describes the random matrix theory we will be using, and applies it to the distinguishability of random quantum states. Section 5 gives the application to the oracle identification problem, and the paper closes with some discussion in section 6.


\section{Bounds on the distinguishability of quantum states}

We consider an ensemble $\mathcal{E}$ containing $n$ $d$-dimensional pure states $\ket{\psi_i}$ with their a priori probabilities $p_i$. We will use $\{\ket{\psi'_i}\}$ to denote the set containing the same states, renormalised to reflect their probabilities (i.e. $\ket{\psi'_i} = \sqrt{p_i} \ket{\psi_i}$). Given an unknown state $\ket{\psi_?}$, picked in accordance with these probabilities, the quantity we are interested in is the average probability of success for a given generalised measurement to distinguish which state we were given. For a measurement $M$ (given by a set of positive operators $\{M_i\}$ summing to the identity), let this probability be denoted by $P^M(\mathcal{E})$. Then we have

\be P^M(\mathcal{E}) = \sum_i \bracket{\psi'_i}{M_i}{\psi'_i} = \sum_i p_i \bracket{\psi_i}{M_i}{\psi_i} \ee
$M^{opt}(\mathcal{E})$ will denote the measurement with the optimal probability of success, and in an abuse of notation $P^{opt}(\mathcal{E})$ will denote this optimal probability. We call this the optimal probability of distinguishing the states in $\mathcal{E}$.

We use three matrix norms: the 1-norm $\|A\|_1 = \sum_{i,j} |A_{ij}|$, the Euclidean (Frobenius) norm $\|A\|_2 = \sqrt{\sum_{i,j} |A_{ij}|^2}$, and the trace norm $\|A\|_{tr} = \mathrm{tr} \sqrt{A^\dag A} = \sum_i \sigma_i(A)$, where $\sigma_i(A)$ denotes the $i$'th singular value of $A$. We will often use the $d \times n$ {\em state matrix} $S = S(\mathcal{E}) = (\ket{\psi'_1},...,\ket{\psi'_n})$ whose $i$'th column is the state $\ket{\psi'_i}$. Then $G=S^{\dag} S$ gives the $n \times n$ Gram matrix \cite{horn} encoding all the inner products between the renormalised states in $\mathcal{E}$. If $n<d$, $G$ will have $d-n$ zero eigenvalues. Note that every rectangular matrix $M$ with $\|M\|_2=1$ is a state matrix. $\rho$ will represent the density matrix of the ensemble:

\be \rho = \sum_{i=1}^n \ket{\psi'_i}\bra{\psi'_i} \ee
It is well-known \cite{jozsa} that $G$ and $\rho$ have the same non-zero eigenvalues.


\subsection{Use of the ``pretty good measurement''}

We will use a specific measurement to provide bounds on $P^{opt}(\mathcal{E})$, which is ``canonical'' in the sense that it performs reasonably well for any ensemble $\mathcal{E}$. This is the so-called {\em pretty good measurement} (PGM), which was independently identified by several authors (e.g. \cite{hausladen}, \cite{hausladen2}) and has a number of useful properties. It is usually defined as a set of projectors $\{\ket{\nu_i}\bra{\nu_i}\}$ onto ``measurement vectors'' $\ket{\nu_i}$, where $\ket{\nu_i} = \rho^{-1/2} \ket{\psi'_i}$ (the inverse only being taken on the support of $\rho$). However, it may also be defined implicitly, which brings out its ``canonical'' nature.

To this end, consider an arbitrary measurement $M$ for $\mathcal{E}$ that consists of a set of $n$ rank 1 projectors onto unnormalised measurement vectors $\ket{\mu_i}$, where each measurement vector corresponds to a state $\ket{\psi'_i}$ in the ensemble. (In fact, it turns out that the optimal measurement for an ensemble of pure states always falls into this category \cite{eldar}.) The probability of getting measurement outcome $i$ and receiving state $j$ is then $|\ip{\mu_i}{\psi'_j}|^2$, and the overall probability of success of this measurement is $\sum_{i=1}^n |\ip{\mu_i}{\psi'_i}|^2$. We may thus encode all the inner products (and hence the probabilities) in a matrix $P$, where $P_{ij} = \ip{\mu_i}{\psi'_j}$; and rather than looking for an optimal measurement $M$, we can rephrase our task as looking for an optimal matrix $P$ that corresponds to a valid measurement.

We have the following requirement on $P$, from the fact that $M$ must be a valid POVM.

\be (P^{\dag} P)_{ij} = \sum_{k=1}^n \ip{\psi'_i}{\mu_k}\ip{\mu_k}{\psi'_j} = \bra{\psi'_i}\left(\sum_{k=1}^n \ket{\mu_k}\bra{\mu_k}\right)\ket{\psi'_j} = G_{ij} = (S^{\dag} S)_{ij} \ee
A natural way to produce a matrix $P$ that satisfies this condition from any given $S$ is to take $P=\sqrt{G}$, the positive semidefinite square root of $G$. The PGM turns out to be a measurement corresponding to this matrix $P$, for, if $P_{ij} = \ip{\nu_i}{\psi'_j}$, then

\be (P^2)_{ij} = \sum_{k=1}^n \bracket{\psi'_i}{\rho^{-1/2}}{\psi'_k} \bracket{\psi'_k}{\rho^{-1/2}}{\psi'_j} = \bra{\psi'_i} \left( \rho^{-1/2} \sum_{k=1}^n \ket{\psi'_k} \bra{\psi'_k} \rho^{-1/2} \right) \ket{\psi'_j} = G_{ij} \ee
The probability of success for the PGM is thus given by $P^{pgm}(\mathcal{E}) = \sum_{i=1}^n (\sqrt{G})_{ii}^2$. Barnum and Knill have proved \cite{barnum} that the PGM has the further property that it is almost optimal in the following sense.

\begin{thm}
\label{pgmNearlyOptimal}
{\bf (Barnum, Knill) \cite{barnum}} $P^{pgm}(\mathcal{E}) \ge P^{opt}(\mathcal{E})^2$.
\end{thm}
So there is the overall relationship $P^{opt}(\mathcal{E})^2 \le P^{pgm}(\mathcal{E}) \le P^{opt}(\mathcal{E})$. For completeness, we include (in Appendix \ref{pgmUpperBoundAppendix}) a simplified proof of Barnum and Knill's result in the case of pure states.


\subsection{Bounds from the pairwise inner products}
\label{innerProductBound}

A set of states that are pairwise almost orthogonal are pairwise almost distinguishable. It thus seems intuitively clear that, given such a set, the probability of success in distinguishing one state from {\em all} the others must also be high. However, this intuition is wrong. This was noted by Jozsa and Schlienz \cite{jozsa}, who showed that the inner products of an ensemble of states may all be reduced, while simultaneously reducing the von Neumann entropy of the ensemble (which gives a measure of overall distinguishability). This effect also manifests itself in quantum fingerprinting \cite{buhrman}. Here, $d$-dimensional states are ``compressed'' to $\log d$-dimensional ``fingerprint'' states that can be distinguished pairwise. However, given such a fingerprint the corresponding original state may not be identified, as this would violate Holevo's theorem \cite{holevo}.

Nevertheless, for certain ensembles the pairwise inner products can give a good lower bound on the overall distinguishability, as noted by several authors \cite{hausladen,barnum}. In this section, we derive such a bound. Our approach is based on that of Hausladen et al. \cite{hausladen}, who found a parabola forming a lower bound on the square root function, which is useful because of the following lemma.

\begin{lem}
If the function $\sqrt{x}$ is bounded below by $f(x) = ax + bx^2$ for $x \ge 0$, then $(\sqrt{G})_{ii} \ge a G_{ii} + b \sum_{j=1}^n |G_{ij}|^2$.
\end{lem}

\begin{proof}
$G$ is a positive semidefinite matrix and thus may be diagonalised: $G = UDU^{\dag}$, where $D = diag(\{\lambda_i\})$ and $U = (u_{ij})$ is unitary. Working out the matrix algebra shows that $(\sqrt{G})_{ii} = \sum_{k=1}^n \sqrt{\lambda_k} |u_{ik}|^2$, so $(\sqrt{G})_{ii} \ge \sum_{k=1}^n f(\lambda_k) |u_{ik}|^2 = f(G)_{ii}$. But $f(G)_{ii} = (aG + bG^2)_{ii} = a G_{ii} + b \sum_{j=1}^n G_{ij} G_{ji} = a G_{ii} + b \sum_{j=1}^n |G_{ij}|^2$.
\end{proof}
Our goal will be to find $a$ and $b$ to parametrise $f$ such that $a G_{ii} + b \sum_{j=1}^n |G_{ij}|^2$ is maximised. It is clear that, for this to be maximised, $f(r)$ must equal $\sqrt{r}$ for some $r$ (or we could just increase $a$ or $b$). So we will pick $a$ and $b$ such that $f(r) = \sqrt{r}$ and $f'(r) = \frac{1}{2 \sqrt{r}}$ (i.e.\ the curves are tangent at this point). This leads to the simultaneous equations
\be ar + br^2 = \sqrt{r} \mbox{,~} a + 2br = \frac{1}{2 \sqrt{r}} \ee
Solving for $a$ and $b$ gives the optimal values

\be a = \frac{3}{2\sqrt{r}} \mbox{,~} b = -\frac{1}{2r^{3/2}} \ee
To see that $f(x)$ actually is a lower bound for $\sqrt{x}$ for any positive value of $r$ (with these values for $a$ and $b$), note that the only solutions to the related equation $f(x)^2=x$ are $x=0$, $x=r$, or $x=4r$. As $f(4r)$ is negative, we have that $f(x)=\sqrt{x}$ if and only if $x=0$ or $x=r$. So the only remaining possibility is that $f(x) > \sqrt{x}$ for all $0 < x < r$. Plugging in a suitable value of $x$ (e.g.\ $r/2$) shows that this is not the case. The expression $a G_{ii} + b \sum_{j=1}^n |G_{ij}|^2$ may now be expressed solely in terms of $r$. Optimising this for $r$ gives that the maximum is found at the point
\be r = \frac{\sum_{j=1}^n |G_{ij}|^2}{G_{ii}} \ee
Returning to the original inequality, we have
\be (\sqrt{G})_{ii} \ge \frac{3}{2\sqrt{r}} G_{ii} - \frac{1}{2r^{3/2}} \sum_{j=1}^n |G_{ij}|^2 \Rightarrow (\sqrt{G})_{ii}^2 \ge \frac{G_{ii}^3}{\sum_{j=1}^n |G_{ij}|^2} \ee
We thus have the following bound on the probability of distinguishing the states in $\mathcal{E}$.
\be \label{ipBound} P^{pgm}(\mathcal{E}) \ge \sum_{i=1}^n \frac{\ip{\psi'_i}{\psi'_i}^3}{\sum_{j=1}^n |\ip{\psi'_i}{\psi'_j}|^2} = \sum_{i=1}^n \frac{p_i^2}{\sum_{j=1}^n p_j |\ip{\psi_i}{\psi_j}|^2} \ee
If all the states have equal a priori probabilities, the bound simplifies further to
\be P^{pgm}(\mathcal{E}) \ge \frac{1}{n} \sum_{i=1}^n \frac{1}{\sum_{j=1}^n |\ip{\psi_i}{\psi_j}|^2} \ee
Unlike previous bounds obtained by other authors for the probability of success of the PGM \cite{hausladen,barnum}, the bound (\ref{ipBound}) is always positive and greater than or equal to $\sum_{i=1}^n p_i^2$, thus showing that the PGM always does at least as well as the ``non-measurement'' of guessing which state was received in accordance with their a priori probabilities.


\subsection{Bounds from eigenvalues}
\label{eigenvalueBound}

The eigenvalues of a Hermitian matrix are closely related to its diagonal elements; indeed, the former majorises the latter \cite{horn}. With this in mind, we look for a bound on the unknown diagonal elements of $\sqrt{G}$ in terms of the known eigenvalues $\{\lambda_i\}$ of $G$.

\begin{lem}
\label{distinguish}
$P^{pgm}(\mathcal{E}) \ge \frac{1}{n} \left(\sum_{i=1}^n \sqrt{\lambda_i} \right)^2 = \frac{1}{n} \|S\|_{tr}^2$.
\end{lem}

\begin{proof}
By the fact that the trace of a matrix is the sum of its eigenvalues, we have

\bea
&& \sum_{i=1}^n (\sqrt{G})_{ii} = \sum_{i=1}^n \sqrt{\lambda_i} \\
&\Rightarrow& \left(\sum_{i=1}^n (\sqrt{G})_{ii}\right)^2 = \left(\sum_{i=1}^n \sqrt{\lambda_i}\right)^2 \\
\label{ineq} &\Rightarrow& n \sum_{i=1}^n (\sqrt{G})_{ii}^2 \ge \left(\sum_{i=1}^n \sqrt{\lambda_i}\right)^2 \\
&\Rightarrow& P^{pgm}(\mathcal{E}) \ge \frac{1}{n} \left(\sum_{i=1}^n \sqrt{\lambda_i} \right)^2
\eea
where in (\ref{ineq}) we used a Cauchy-Schwarz inequality, showing that equality can only be attained in step (\ref{ineq}) when all the $(\sqrt{G})_{ii}$ are equal.
\end{proof}

Interestingly, this bound is the same as the fidelity of $G$ with the maximally mixed state $I/n$, where the fidelity $F(\rho,\sigma)$ is defined as $\left(\mbox{tr~}\sqrt{\rho^{1/2}\,\sigma\,\rho^{1/2}}\right)^2$ \cite{nielsen}.

It is worth noting that no upper bound on the success probability in terms of the eigenvalues alone can be found, for the following reason. Any set of eigenvalues $\{\lambda_i\}$ summing to 1 can give rise to a Gram matrix $G$ where $G_{ii} = \lambda_i$, and $G_{ij} = 0$ (for $i \neq j$). Such matrices correspond to an ensemble $\mathcal{E}$ of perfectly distinguishable states where $P^{pgm}(\mathcal{E}) = 1$. As future work, it would be interesting to determine whether an upper bound (or an improved lower bound) could be produced by considering the diagonal entries of $G$ as well as its eigenvalues.


\subsection{Distinguishing mixed states}

It is natural to ask to what extent these lower bounds hold for the generalised problem of distinguishing an ensemble $\mathcal{E}$ consisting of mixed states $\{\rho_i\}$. The following lemma allows the problem to be related to that of distinguishing pure states.

\begin{lem}
\label{mixedToPure}
Let $\mathcal{E}$ be an ensemble of $n$ $d$-dimensional mixed states $\{\rho_i\}$ with a priori probabilities $\{p_i\}$, and having spectral decompositions $\rho_i = \sum_{k=1}^d \lambda_{ik} \ket{v_{ik}}\bra{v_{ik}}$. Let $\mathcal{F}$ be an ensemble of the $nd$ pure states given by the eigenvectors $\{\ket{v_{ik}}\}$ with a priori probabilities $\{p_i \lambda_{ik}\}$. Then $P^{pgm}(\mathcal{E}) \ge P^{pgm}(\mathcal{F})$.
\end{lem}

\begin{proof}
For mixed states, the PGM is defined by the following measurement operators $\{M_i\}$:

\be M_i = \rho^{-1/2} \rho'_i \rho^{-1/2},\, \mbox{~where~} \rho'_i = p_i \rho_i \mbox{~and~} \rho = \sum_{i=1}^n \rho'_i \ee
So the probability of success can be bounded as follows, where we use the renormalised eigenvectors $\ket{v'_{ik}} = \sqrt{p_i} \sqrt{\lambda_{ik}} \ket{v_{ik}}$.

\bea
P^{pgm}(\mathcal{E}) &=& \sum_{i=1}^n \mathrm{tr}\left(\rho^{-1/2} \rho'_i \rho^{-1/2} \rho'_i\right) \\
&=& \sum_{i=1}^n \mathrm{tr}\left(\rho^{-1/2} \left(\sum_{k=1}^d \ket{v'_{ik}}\bra{v'_{ik}}\right) \rho^{-1/2} \left(\sum_{l=1}^d \ket{v'_{il}}\bra{v'_{il}}\right)\right) \\
&=& \sum_{i=1}^n \sum_{k,l=1}^d \mathrm{tr}\left(\rho^{-1/2}\ket{v'_{ik}}\bra{v'_{ik}} \rho^{-1/2}\ket{v'_{il}}\bra{v'_{il}}\right) \\
&=& \sum_{i=1}^n \sum_{k,l=1}^d |\bracket{v'_{ik}}{\rho^{-1/2}}{v'_{il}}|^2 \ge \sum_{i=1}^n \sum_{k=1}^d |\bracket{v'_{ik}}{\rho^{-1/2}}{v'_{ik}}|^2 = P^{pgm}(\mathcal{F})
\eea
\end{proof}

Therefore, if the eigenvalues and eigenvectors of the states $\{\rho_i\}$ are known, the lower bounds given previously may be applied. If not, a weaker lower bound based only on the pairwise fidelities of the states may be given (where, as before, we set $F(\rho,\sigma)=\left(\mbox{tr~}\sqrt{\rho^{1/2}\,\sigma\,\rho^{1/2}}\right)^2$).

\begin{thm}
Let $\mathcal{E}$ be an ensemble of $n$ $d$-dimensional mixed states $\{\rho_i\}$ with a priori probabilities $\{p_i\}$. Then

\be P^{pgm}(\mathcal{E}) \ge \sum_{i=1}^n \frac{p_i^2\,\mathrm{tr}(\rho_i^2)}{\sum_{j=1}^n p_j F(\rho_i,\rho_j)} \ee
\end{thm}

\begin{proof}
From the bound ($\ref{ipBound}$) and Lemma \ref{mixedToPure}, we have
\bea
P^{pgm}(\mathcal{E}) &\ge& \sum_{i=1}^n \sum_{k=1}^d \frac{p_i^2 \lambda_{ik}^2}{\sum_{j=1}^n \sum_{l=1}^d p_j \lambda_{jl} |\ip{v_{ik}}{v_{jl}}|^2}\\
&=& \sum_{i=1}^n \sum_{k=1}^d \frac{p_i^2 \lambda_{ik}^2}{\sum_{j=1}^n p_j \bra{v_{ik}}\left(\sum_{l=1}^d \lambda_{jl} \ket{v_{jl}} \bra{v_{jl}}\right) \ket{v_{ik}}} \\
&=& \sum_{i=1}^n \sum_{k=1}^d \frac{p_i^2 \lambda_{ik}^2}{\sum_{j=1}^n p_j \bracket{v_{ik}}{\rho_j}{v_{ik}}} \\
&\ge& \sum_{i=1}^n \sum_{k=1}^d \frac{p_i^2 \lambda_{ik}^2}{\sum_{j=1}^n p_j F(\rho_i,\rho_j)} = \sum_{i=1}^n \frac{p_i^2\,\mathrm{tr}(\rho_i^2)}{\sum_{j=1}^n p_j F(\rho_i,\rho_j)}
\eea
\end{proof}

This bound gets progressively worse as the states in $\mathcal{E}$ get more mixed. One might expect the following lower bound to hold for mixed states, as it is the obvious extension of the bound (\ref{ipBound}) for pure states, but interestingly it does not.

\be P^{pgm}(\mathcal{E}) \ngeq \sum_{i=1}^n \frac{p_i^2}{\sum_{j=1}^n p_j F(\rho_i,\rho_j)} \ee
A simple counterexample is given by the equiprobable ensemble consisting of the following two three-dimensional states.

\be \rho_1 = \sm{\frac{1}{2} & 0 & 0 \\ 0 & \frac{1}{2} & 0 \\ 0 & 0 & 0},\,\rho_2 = \sm{\frac{1}{2} & 0 & 0 \\ 0 & 0 & 0 \\ 0 & 0 & \frac{1}{2}} \ee


\section{The distinguishability of states with constant inner product}

An illustrative case to apply these bounds to is that of equiprobable states where the pairwise inner products are all equal, so the states are all equally distinguishable from each other. Consider an ensemble $\mathcal{E}$ with Gram matrix $G$, where $G_{ii} = 1/n$ and $G_{ij}=p/n$ for $i \neq j$ (and $p$ is a positive real constant). In this case, the inner product bound of section \ref{innerProductBound} gives the bound

\be P^{pgm}(\mathcal{E}) \ge \frac{1}{1+p^2(n-1)} = O(1/n) \ee
The eigenvalue bound, however, gives much better results. The symmetry of $G$ shows immediately that it has an eigenvector $(1,1,...,1)$; the corresponding eigenvalue is $\lambda_1 = p+(1-p)/n$. The set of eigenvectors may be completed by taking any $n-1$ vectors orthogonal to $(1,1,...,1)$, which will be eigenvectors with eigenvalues $\lambda_{2...n}=(1-p)/n$. We therefore have

\bea
\label{pgmTight} P^{pgm}(\mathcal{E}) &\ge& \frac{1}{n} \left(\sqrt{p + \frac{1-p}{n}} + (n-1) \sqrt{\frac{1-p}{n}} \right)^2 \\
&\ge& \frac{1}{n} \left((n-1)^2\frac{(1-p)}{n}\right) \ge (1-p)-\frac{2(1-p)}{n}
\eea
so the probability of distinguishing these states approaches a constant as $n\rightarrow \infty$. In fact, one can show that inequality (\ref{pgmTight}) is actually an equality giving the precise probability of success $P^{pgm}(\mathcal{E})$ (this follows from showing that the diagonal entries of $\sqrt{G}$ are all equal).

Such an ensemble therefore provides a kind of converse to the ensemble of states used in quantum fingerprinting \cite{buhrman}: in this case, no matter how many states there are in the ensemble, their joint distinguishability is of the same order as their pairwise distinguishability. We will see below that this behaviour is not typical; however, it is perhaps not surprising, because $\mathcal{E}$ can only be realised in $n$ dimensions. To see this, note that $G$ is non-singular, so the states in $\mathcal{E}$ must be linearly independent.


\section{The distinguishability of random quantum states}

We will use Lemma \ref{distinguish} and some results from the theory of random matrices to put a lower bound on the probability of distinguishing random quantum states. The expected value of this lower bound will be obtained for a quite general notion of ``randomness'', but in order to get measure concentration results we will specialise to states distributed uniformly at random (according to the Haar measure). The results hold in the asymptotic regime where the number of states $n$ and the dimension $d$ approach a constant ratio.


\subsection{A little random matrix theory}

In this section, we will calculate the expected value of the trace norm of a random matrix. The distribution of the trace norm (i.e.\ the sum of singular values) of a matrix $M$ is clearly related to that of the eigenvalues of the matrix $MM^\dag$, which is known to statisticians as a (complex) {\em Wishart matrix}. The distribution of the eigenvalues of a Wishart matrix is given by the Mar\v{c}enko-Pastur law \cite{marchenko}, which is stated in the form we need in \cite{bai}.

\begin{thm}
\label{mplaw}
{\bf (Mar\v{c}enko/Pastur law) \cite{marchenko}}\\
Let $R_r$ be a family of $d \times n$ matrices with $n \ge d$ and $d/n \rightarrow r \in (0,1]$ as $n,d \rightarrow \infty$, where the entries of $R_r$ are i.i.d.\ complex random variables with mean 0 and variance 1. Then, as $n,d \rightarrow \infty$, the eigenvalues of the rescaled matrix $\frac{1}{n} R_r R_r^\dag$ tend to a limiting distribution with density

\be \label{eigenPdf} p_r(x) = \frac{\sqrt{(x-A^2)(B^2-x)}}{2 \pi r x} \ee
for $A^2 \le x \le B^2$ (where $A = 1-\sqrt{r}$, $B = 1+\sqrt{r}$), and density 0 elsewhere.
\end{thm}
We will translate this to a similar statement about the singular values of $R_r$. The following lemma is straightforward.

\begin{lem}
\label{singularValueDistribution}
Let $R_r$ be a family of $d \times n$ matrices with $k/m \rightarrow r \in (0,1]$ as $n,d \rightarrow \infty$, where $k=\min(n,d)$ and $m=\max(n,d)$, and the entries of $R_r$ are i.i.d.\ complex random variables with mean 0 and variance 1. Then, as $n,d \rightarrow \infty$, the singular values of $R_r/\sqrt{m}$ tend to a limiting distribution with density

\be \label{singPdf} p_r(y) = \frac{\sqrt{(y^2-A^2)(B^2-y^2)}}{\pi r y} \ee
for $A \le y \le B$ (where $A = 1-\sqrt{r}$, $B = 1+\sqrt{r}$), and density 0 elsewhere.
\end{lem}

\begin{proof}
The lemma follows from Theorem \ref{mplaw} for $n \ge d$ by substituting $y=\sqrt{x}$. For $n \le d$, note that the singular values of $R$ are the same as those of $R^T$, so the roles of $n$ and $d$ need merely be interchanged.
\end{proof}

\begin{lem}
Let $R_r$ be a family of $d \times n$ matrices with $k/m \rightarrow r \in (0,1]$ as $n,d \rightarrow \infty$, where $k=\min(n,d)$ and $m=\max(n,d)$, and the entries of $R_r$ are i.i.d.\ complex random variables with mean 0 and variance 1. Then, as $n,d \rightarrow \infty$, the expected trace norm of $R_r$ is

\be \label{ellipticInteg} \E(\|R_r\|_{tr}) = \frac{m^{3/2}}{\pi} \int_A^B \sqrt{(y^2-A^2)(B^2-y^2)}\,dy \ee
where $A = 1-\sqrt{r}$, $B = 1+\sqrt{r}$.
\end{lem}

\begin{proof}
With probability 1, $R_r$ will have $k$ non-zero singular values. Let $\sigma_i(R_r)$ denote the value of the $i$'th (unsorted) singular value of $R_r$, for arbitrary $i$ between 1 and $k$. We have

\be \E(\|R_r\|_{tr}) = (k \sqrt{m})\,\E(\sigma_i(R_r/\sqrt{m})) = k\sqrt{m} \int_A^B y\,p_r(y)\,dy \ee
and using Lemma \ref{singularValueDistribution} gives the desired result.
\end{proof}
This turns out to be an elliptic integral which cannot be expressed in terms of elementary functions \cite{gradshteyn}. However, it is possible to produce a good lower bound, which is tight in the case $r=1$:

\begin{lem}
\label{expectedTraceNorm}
\be \E(\|R_r\|_{tr}) \ge k \sqrt{m} \sqrt{1-r\left(1-\frac{64}{9\pi^2}\right)} \ee
with equality when $r=1$.
\end{lem}

\begin{proof}
See Appendix \ref{traceNormAppendix}.
\end{proof}


\subsection{Random quantum states}

Knowing the expected value of the trace norm immediately allows us to say something about the expected distinguishability of an ensemble of random quantum states, for a quite general notion of randomness.

\begin{thm}
\label{randomStates}
Let $\mathcal{E}$ be an ensemble of $n$ equiprobable $d$-dimensional quantum states $\{\ket{\psi_i}\}$ with $n/d \rightarrow r \in (0,\infty)$ as $n,d \rightarrow \infty$, and let the components of $\ket{\psi_i}$ in some basis be i.i.d. complex random variables with mean 0 and variance $1/d$. Then

\be
\E(P^{pgm}(\mathcal{E})) \ge \left\{ \begin{array}{ll}
\frac{1}{r}\left(1-\frac{1}{r}\left(1-\frac{64}{9 \pi^2}\right)\right) & \mbox{if~} n \ge d\\
1-r\left(1-\frac{64}{9 \pi^2}\right) & \mbox{otherwise} \\
\end{array}\right.
\ee
and in particular $\E(P^{pgm}(\mathcal{E})) > 0.720$ when $n \le d$.
\end{thm}

\begin{proof}
The matrix $R=\sqrt{nd}\,S(\mathcal{E})$ fulfils the criteria for the Mar\v{c}enko-Pastur law (\ref{mplaw}), as its entries are complex random variables with mean 0 and variance 1. We therefore have

\be
\E(P^{pgm}(\mathcal{E})) \ge \E\left(\frac{1}{n} \|S(\mathcal{E})\|_{tr}^2\right) \ge \frac{1}{n} \E(\|S(\mathcal{E})\|_{tr})^2 = \frac{1}{n^2 d} \E(\|R\|_{tr})^2
\ee
and plugging in the lower bound on the expected trace norm of $R$ from Lemma \ref{expectedTraceNorm} gives the required result.
\end{proof}

We can immediately apply this result to the distinguishability of random quantum states uniformly distributed on the complex unit sphere in $d$ dimensions. A uniformly random quantum state may be produced by creating a vector $v$, each of whose components are complex Gaussians (say $v_i \sim \tilde{N}(0,1/d)$), and normalising the result. By the law of large numbers, as $d \rightarrow \infty$, the norm of the resulting vector will approach $1$, so the normalisation step becomes unnecessary. (This can be formalised and is known as Poincar\'e's lemma \cite{ledoux2}.) Therefore, an ensemble of uniformly random states meets the criteria for Theorem \ref{randomStates}, so we can lower bound its expected distinguishability.

In fact, in this case, we may exploit the concentration of measure effects characteristic of high-dimensional spaces to show that for high $d$ {\em almost all} ensembles of $n\le d$ states are quite distinguishable. As with the recent paper \cite{popescu}, our tool will be Levy's Lemma \cite{ledoux}:

\begin{lem}
{\bf (Levy's Lemma) \cite{ledoux}}\\
Given a function $f: \mathbb{S}^d \mapsto \R$ defined on the $d$-dimensional real hypersphere $\mathbb{S}^d$, and a point $p$ on the hypersphere chosen uniformly at random,

\be \Pr[|f(p)-\E(f)|\ge \epsilon] \le 2 \exp\left(\frac{-2 C(d+1)\epsilon^2}{\eta^2}\right) \ee
where $\eta$ is the Lipschitz constant of $f$, $\eta = \sup_{x,y} |f(x)-f(y)|/\|x-y\|_2$, and $C$ is a positive constant that may be taken to be $1/(18\pi^3)$.
\end{lem}
This is useful for us because a state matrix is precisely such a point on a hypersphere:

\begin{lem}
Let $\mathcal{E}$ be an ensemble of $n$ equiprobable $d$-dimensional quantum states picked uniformly at random. Then, for large $d$, the state matrix $S(\mathcal{E})$ defines a point picked uniformly at random on the sphere in $nd$ complex dimensions (equivalently, the real sphere $\mathbb{S}^{2nd-1}$ in $2nd$ dimensions).
\end{lem}

\begin{proof}
As noted previously, by the properties of quantum states distributed uniformly at random, for high $d$ the elements of $S(\mathcal{E})$ will be complex Gaussians with mean 0 and variance $1/nd$. The lemma follows.
\end{proof}

\begin{lem}
\label{lipschitz}
Let $S$ be an $n \times d$ matrix with $\|S\|_2=1$, and define $f(S) = \frac{1}{n}\|S\|_{tr}^2$. Then the Lipschitz constant $\eta$ of $f$ satisfies $\eta \le 2$.
\end{lem}

\begin{proof}
See Appendix \ref{lipschitzAppendix}.
\end{proof}
Plugging this function $f$ and this value of $\eta$ into Levy's Lemma gives the following theorem.

\begin{figure}[htp]
\begin{center}
\subfigure[$0\le r \le 2$]{
\includegraphics[clip,width=70mm,viewport=40 60 580 430]{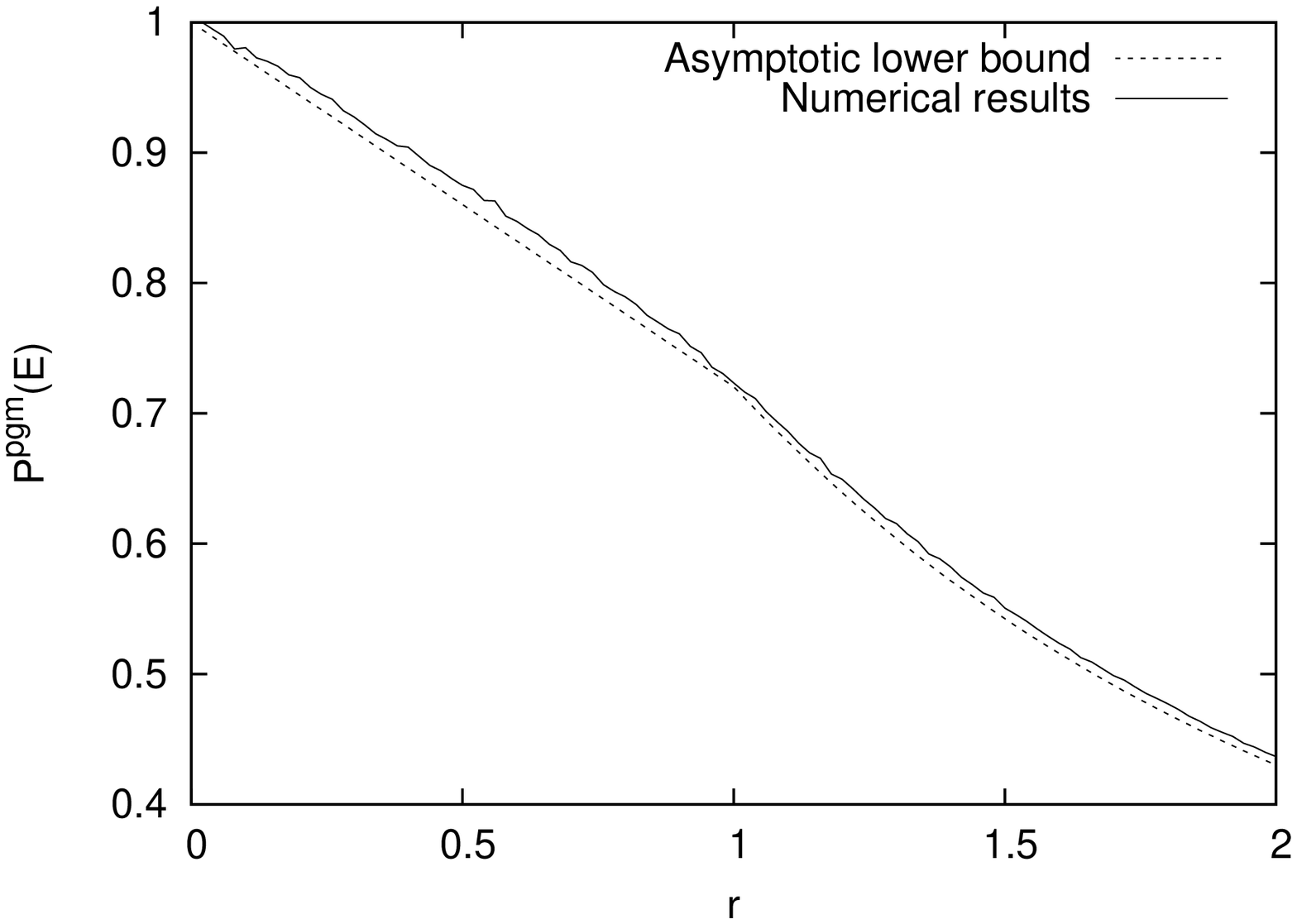}}
\subfigure[$0\le r \le 10$]{
\includegraphics[clip,width=70mm,viewport=40 60 580 430]{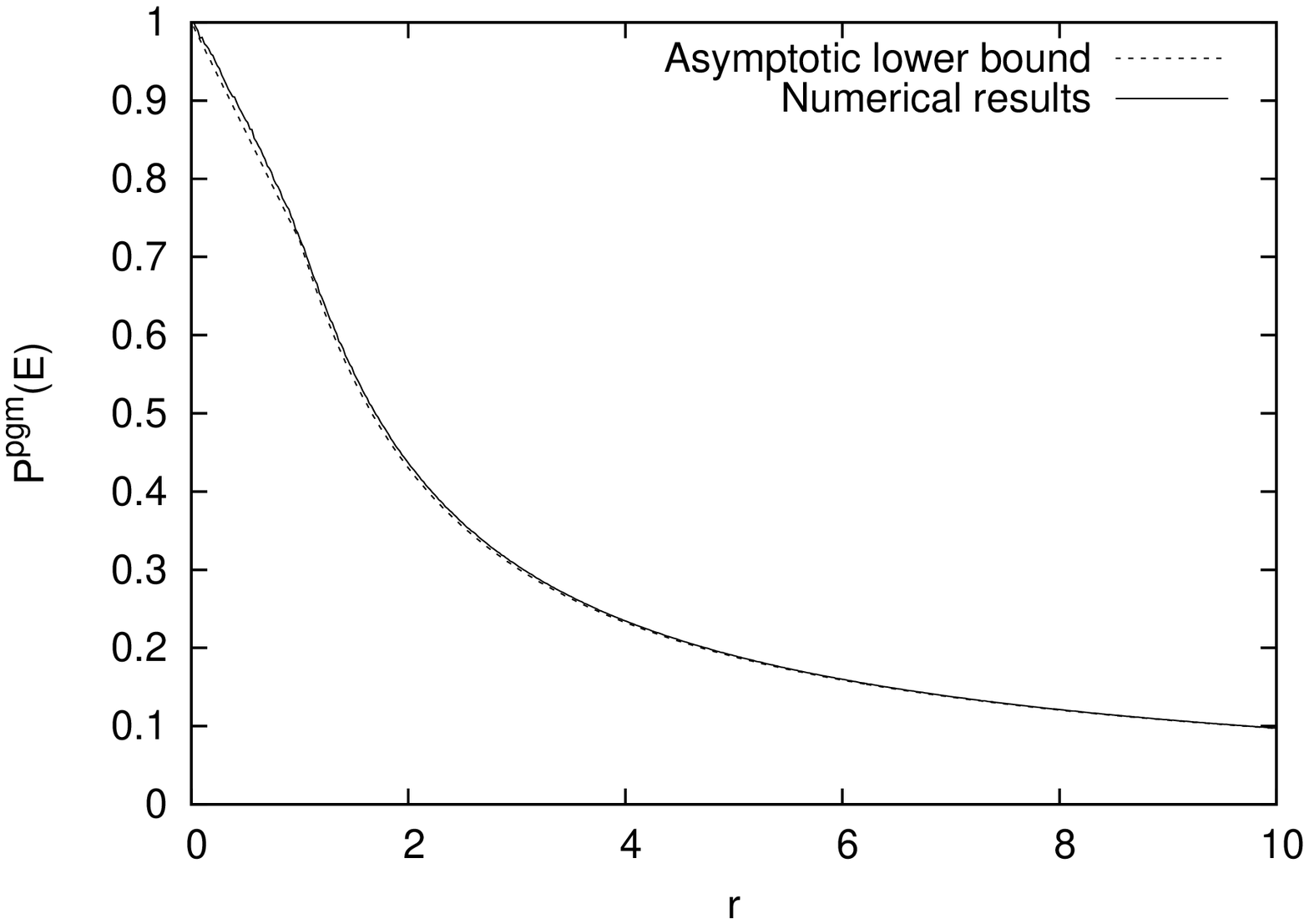}}
\caption{Asymptotic bound on $P^{pgm}(\mathcal{E})$ vs.\ numerical results (averaged over 10 runs) for ensembles of $n=50r$ 50-dimensional uniformly random states.}
\label{fig1}
\end{center}
\end{figure}

\begin{thm}
Let $\mathcal{E}$ be an ensemble of $n$ $d$-dimensional quantum states picked uniformly at random. Set $p = \E(P^{pgm}(\mathcal{E})) = \frac{1}{r}\left(1-\frac{1}{r}\left(1-\frac{64}{9 \pi^2}\right)\right)$ if $n \ge d$, and $p = 1-r\left(1-\frac{64}{9 \pi^2}\right)$ otherwise. Then

\be \Pr[P^{pgm}(\mathcal{E}) \le p-\epsilon] \le 2 \exp\left(\frac{-C(2nd+1)\epsilon^2}{2}\right) \ee
where $C=1/(18\pi^3)$.
\end{thm}
Figure \ref{fig1} shows numerical evidence that ensembles $\mathcal{E}$ of quantum states picked uniformly at random appear to have a value of $P^{pgm}(\mathcal{E})$ close to this lower bound, even when the states are (relatively) low-dimensional.

\section{Application to oracle identification}

The {\em oracle identification problem} may be defined as follows \cite{ambainis}. Given an unknown $n$-bit Boolean function $f: \{0,1\}^n \mapsto \{0,1\}$ (the {\em oracle}), picked uniformly at random from a known set $F$ of functions, identify $f$ with the minimum number of uses of $f$. Set $N=|F|$ and $D=2^n$. Clearly, classical computation cannot identify $f$ with fewer than $\log_2 N$ queries in the worst case (as each query may reduce the search space by at most half). However, quantum computation can sometimes do better. On a quantum computer, we can encode the oracle as an $n$ qubit unitary operator $U_f$, defined by the action $U_f\ket{x} \mapsto (-1)^{f(x)} \ket{x}$. Now if the uniform superposition $\frac{1}{2^{n-1}}\sum_{x=0}^{2^n-1} \ket{x}$ is input to the oracle, the following {\em oracle state} will be be produced:

\be \ket{\psi_f} = \frac{1}{2^{n-1}}\sum_{x=0}^{2^n-1} (-1)^{f(x)} \ket{x} \ee
In some cases, a single quantum query to $U_f$ may be enough to identify $f$ with certainty. This will be the case if $\ip{\psi_f}{\psi_g}=0$ for all $f\neq g$ (although this is not a necessary condition). The satisfaction of this orthogonality condition may be expected to be a rare event, and is certainly impossible when $N>D$. However, if we are content with a small probability of error, the situation is better: we will show here that, in particular, {\em almost all} sets of $N=D$ oracles may be distinguished almost certainly in a constant number of quantum queries.

The oracle identification problem was introduced and studied by Ambainis et al \cite{ambainis}, who (among other results) developed a hybrid quantum-classical algorithm for the random oracle case with which we concern ourselves here. However, the upper bound they obtained in the case where $N=D$ is only $O(\log_2 N)$ queries, which is no better than classical computation.

\begin{lem}
Let $\mathcal{E}$ be an ensemble of $N$ $D$-dimensional oracle states corresponding to Boolean functions picked uniformly at random (call these {\em random oracle states}). Then the rescaled state matrix $\sqrt{ND}\,S(\mathcal{E})$ defines a point picked uniformly at random on the $ND$-dimensional hypercube $\{-1,1\}^{ND}$.
\end{lem}

\begin{proof}
Each component of each state will be $\pm1/\sqrt{ND}$, with equal probability of each.
\end{proof}
$\sqrt{ND}\,S(\mathcal{E})$ therefore meets the required conditions for the Mar\v{c}enko-Pastur law (\ref{mplaw}), so we may say immediately

\begin{lem}
Let $\mathcal{E}$ be an ensemble of $N$ $D$-dimensional random oracle states, and set $r=N/D$. Then

\be
\E(P^{pgm}(\mathcal{E})) \ge \left\{ \begin{array}{ll}
\frac{1}{r}\left(1-\frac{1}{r}\left(1-\frac{64}{9 \pi^2}\right)\right) & \mbox{if~} N \ge D\\
1-r\left(1-\frac{64}{9 \pi^2}\right) & \mbox{otherwise} \\
\end{array}\right.
\ee
and in particular $\E(P^{pgm}(\mathcal{E})) \ge 0.720$ when $N \le D$.
\end{lem}

Like the sphere, the high-dimensional hypercube exhibits the concentration of measure phenomenon, and we can write down a similar result to Levy's Lemma \cite{ledoux}:

\begin{lem}
\label{levycube}
{\bf (Concentration of measure on the cube) \cite{ledoux}}\\
Given a function $f: \{-1,1\}^d \mapsto \R$ defined on a $d$-dimensional hypercube, and a point $p$ on the hypercube chosen uniformly at random,

\be \Pr[|f(p)-\E(f)|\ge \epsilon] \le 2 \exp\left(\frac{-2\epsilon^2}{d\eta^2}\right) \ee
where $\eta$ is the Lipschitz constant of $f$ with respect to the Hamming distance, $\eta = \sup_{x,y} |f(x)-f(y)|/d(x,y)$.
\end{lem}

\begin{lem}
\label{lipschitzCube}
Let $H$ be a point on the $nd$-dimensional hypercube written down as an $n \times d$ $\{-1,1\}$-matrix, and let $f(H) = \frac{1}{n^2 d}\|H\|_{tr}^2$. Then the Lipschitz constant $\eta$ of $f$ satisfies $\eta \le 4/nd$.
\end{lem}

\begin{proof}
See Appendix \ref{lipschitzAppendix}.
\end{proof}

Plugging this value of $\eta$ into Lemma \ref{levycube} gives

\begin{thm}
Let $\mathcal{E}$ be an ensemble of $N$ $D$-dimensional random oracle states. Set $p = \E(P^{pgm}(\mathcal{E})) = \frac{1}{r}\left(1-\frac{1}{r}\left(1-\frac{64}{9 \pi^2}\right)\right)$ if $N \ge D$, and $p = 1-r\left(1-\frac{64}{9 \pi^2}\right)$ otherwise, where $r=N/D$. Then

\be \Pr[P^{pgm}(\mathcal{E}) \le p-\epsilon] \le 2 \exp\left(\frac{-2ND\epsilon^2}{16}\right) \ee
\end{thm}
and we have our desired result: with 1 query, all but an exponentially small fraction of the possible sets of $N$ $N$-dimensional random oracle states may be distinguished with a constant probability bounded away from $1/2$ (in fact, to get a probability of success greater than $1/2$, we may take $r=N/D$ to be as high as $\sim 1.66$). A constant number of repetitions allows this probability to be boosted to be arbitrarily high.


\section{Discussion}

This work can be seen as part of an overall programme of understanding the behaviour of random quantum states \cite{woottersRandom,popescu,winter,sommers}.

There is a fundamental correspondence between the mixed state obtained from an equal mixture of uniformly random pure states, and that produced by starting with a larger system in a uniformly random pure state, and tracing out part of the system. Consider a $d$-dimensional state

\be \rho_{n,d} = \frac{1}{n} \sum_{i=1}^n \ket{\psi_i}\bra{\psi_i} \ee
where each state in the set $\mathcal{E}=\{\ket{\psi_i}\}$ is picked uniformly at random. We can think of $\rho_{n,d}$ as being produced from the following $dn$-dimensional state (which we consider to live in a Hilbert space $\mathcal{H}_d \otimes \mathcal{H}_n$) by tracing out the second subsystem:

\be \ket{\upsilon} = \frac{1}{\sqrt{n}} \sum_{k=0}^{n-1} \ket{\upsilon_k} \ket{k} = \frac{1}{\sqrt{n}} \sum_{k=0}^{n-1} \sum_{l=0}^{d-1} \alpha_{kl} \ket{l} \ket{k} \ee
for some coefficients $\alpha_{kl}$. As mentioned previously, the $\alpha_{kl}$ will be approximately normally distributed as $\tilde{N}(0,1/d)$. So, because of the normalisation factor at the front of the sum, the overall state $\ket{\upsilon}$ has coefficients which are normally distributed and scaled as $\tilde{N}(0,1/dn)$. Therefore, this state is picked from the uniform distribution on the unit sphere in $\C^{dn}$. Popescu, Short and Winter \cite{popescu} obtained an upper bound on the expected trace distance of such a state $\rho_{n,d}$ from the maximally mixed state $I/d$, and used this to show that for $n \gg d$, $\rho \approx I/d$.

Because the non-zero eigenvalues of the Gram matrix of (rescaled) states in $\mathcal{E}$ are the same as the eigenvalues of $\rho_{n,d}$ \cite{jozsa}, this paper can be seen as obtaining a similar result to \cite{popescu} for the {\em fidelity} of $\rho_{n,d}$ with the maximally mixed state, via quite different methods. However, the bound is tighter for $n$ close to $d$, and the notion of ``randomness'' of the states $\{\ket{\psi_i}\}$ is more general (which is simply a side-effect of relying on the powerful Mar\v{c}enko-Pastur law).


\section*{Acknowledgements}

I would like to thank Richard Jozsa for careful reading of this manuscript, and Aram Harrow and Tony Short for helpful discussions. I would also like to thank Jon Tyson for pointing out an error in Appendix \ref{pgmUpperBoundAppendix}. This work was supported in part by the UK Engineering and Physical Sciences Research Council QIP-IRC grant.

\appendix

\section*{Appendices}


\section{The PGM is close to optimal}
\label{pgmUpperBoundAppendix}

\begin{pgmNearlyOptimalThm*}
{\bf (Barnum, Knill) \cite{barnum}} $P^{pgm}(\mathcal{E}) \ge P^{opt}(\mathcal{E})^2$.
\end{pgmNearlyOptimalThm*}

\begin{proof}
Consider an arbitrary POVM $R$ consisting of measurement operators $\{R_i\}$, and an arbitrary ensemble $\mathcal{E}$ of renormalised states $\{\ket{\psi'_i}\}$, with a priori probabilities $p_i$, where as before $\ket{\psi'_i} = \sqrt{p_i} \ket{\psi_i}$ and $\rho = \sum_{i=1}^n \ket{\psi'_i}\bra{\psi'_i}$. Assume wlog that $R_i = \ket{\mu_i}\bra{\mu_i}$ for some vectors $\ket{\mu_i}$, as the optimal measurement will always be of this form \cite{eldar}. Then

\bea
P^R(\mathcal{E}) &=& \sum_{i=1}^n \bracket{\psi'_i}{R_i}{\psi'_i} = \sum_{i=1}^n |\ip{\psi'_i}{\mu_i}|^2 = \sum_{i=1}^n |\bracket{\psi'_i}{\rho^{-1/4} \rho^{1/4}}{\mu_i}|^2 \\
&\le& \sum_{i=1}^n \bracket{\psi'_i}{\rho^{-1/2}}{\psi'_i} \bracket{\mu_i}{\rho^{1/2}}{\mu_i} \\
&\le& \sqrt{\left(\sum_{i=1}^n \bracket{\psi'_i}{\rho^{-1/2}}{\psi'_i}^2\right) \left(\sum_{j=1}^n \bracket{\mu_j}{\rho^{1/2}}{\mu_j}^2\right)} \\
&\le& \sqrt{\sum_{i=1}^n \bracket{\psi'_i}{\rho^{-1/2}}{\psi'_i}^2} = \sqrt{P^{pgm}(\mathcal{E})}
\eea
The first and second inequalities are Cauchy-Schwarz inequalities, and the third follows because the vectors $\{\rho^{1/2}\ket{\mu_i}\}$ can easily be seen to define an ensemble with density matrix $\rho$:

\be \sum_{i=1}^n \rho^{1/2}\ket{\mu_i}\bra{\mu_i}\rho^{1/2} = \rho^{1/2}\left(\sum_{i=1}^n \ket{\mu_i}\bra{\mu_i}\right)\rho^{1/2} = \rho \ee
and we therefore have $\sum_{i=1}^n \bracket{\mu_i}{\rho^{1/2}}{\mu_i}^2 \le 1$, as this is the probability of success of the measurement $R$ applied to this ensemble.
\end{proof}


\section{Proof of Lemma \ref{expectedTraceNorm}}
\label{traceNormAppendix}

In this appendix we will prove a lemma which immediately implies Lemma \ref{expectedTraceNorm}. See \cite{gradshteyn} for the facts used about elliptic integrals and hypergeometric series.

\begin{lem}
Let $0 \le r \le 1$ and $A=1-\sqrt{r}$, $B=1+\sqrt{r}$. Then

\be \label{integralIneq} \int_A^B \sqrt{(y^2-A^2)(B^2-y^2)}\,dy \ge r\pi\sqrt{1-r\left(1-\frac{64}{9\pi^2}\right)} \ee
with equality at $r=0$, $r=1$.
\end{lem}

\begin{proof}
We have

\bea
f(r) &=& \int_A^B \sqrt{(y^2-A^2)(B^2-y^2)}\,dy\\
&=& \frac{B}{3} \left( (A^2+B^2) E\left(\frac{\sqrt{B^2-A^2}}{B^2}\right) - 2A^2 K\left(\frac{\sqrt{B^2-A^2}}{B^2}\right) \right)\\
&=& \frac{2(1+\sqrt{r})}{3} \left( (1+r) E\left(\frac{2r^{1/4}}{1+\sqrt{r}}\right) - (1-\sqrt{r})^2 K\left(\frac{2r^{1/4}}{1+\sqrt{r}}\right) \right)
\eea
where $K(r)$ and $E(r)$ are the complete elliptic integrals of the first and second kind, respectively:

\be K(r) = \int_0^1 \frac{dx}{\sqrt{(1-x^2)(1-r^2 x^2)}}\,,\, E(r) = \int_0^1 \frac{\sqrt{1-r^2 x^2}}{\sqrt{1-x^2}} dx \ee
Note that $f(r)$ may be evaluated explicitly for $r=0$ and $r=1$, giving 0 and $8/3$ respectively. Now we may apply a standard change of variables (Landen's transformation) to both elliptic integrals, giving

\bea
\nonumber f(r) &=& \frac{2(1+\sqrt{r})}{3} \left( \frac{1+r}{1+\sqrt{r}}\left( 2 E(\sqrt{r}) - (1-r) K(\sqrt{r}) \right) -(1-\sqrt{r})^2(1+\sqrt{r})K(\sqrt{r})\right) \\
&=& \frac{4}{3} \left( (1+r)E(\sqrt{r})-(1-r)K(\sqrt{r}) \right)
\eea
We now move to the representation of $K(r)$ and $E(r)$ as hypergeometric series, which are defined as follows (using the notation $a^{\bar{n}} = a(a+1)\cdots(a+n-1)$).

\be {}_2F_1(a,b;c;r) = \sum_{n=0}^\infty \frac{a^{\bar{n}}b^{\bar{n}}}{c^{\bar{n}}n!} r^n\ee

\be K(r) = (\pi/2)\,{}_2F_1(1/2,1/2;1;r^2) \,,\, E(r) = (\pi/2)\,{}_2F_1(-1/2,1/2;1;r^2) \ee
This has the advantage that, by a transformation rule due to Gauss, we can rewrite $f(r)$ as a single hypergeometric series.

\bea
f(r) &=& \frac{2\pi}{3} \left( (1+r)\,{}_2F_1(-1/2,1/2;1;r) - (1-r)\,{}_2F_1(1/2,1/2;1;r) \right) \\
&=& \pi r\,{}_2F_1(-1/2,1/2;2;r)
\eea
Returning to the original inequality, our task has been simplified to showing that

\be g(r) = {}_2F_1(-1/2,1/2;2;r)^2 \ge 1-r\left(1-\frac{64}{9\pi^2}\right) \ee
Evaluating $g(r)$ at 0 and 1 makes it clear that this is equivalent to showing that $g(r)$ is concave for $0 \le r \le 1$, which would follow from showing the second derivative $g''(r)$ to be negative in this region. From the rules governing differentiation of hypergeometric series, it is easy to show that

\be g''(r) = \frac{1}{32}\left({}_2F_1(1/2,3/2;3;r)^2 - 2\,{}_2F_1(-1/2,1/2;2;r) {}_2F_1(3/2,5/2;4;r) \right) \ee
The following hypergeometric transformation allows this to be simplified.

\bea
&& {}_2F_1(a,b;c;r) = (1-r)^{c-a-b}{}_2F_1(c-a,c-b;c;r) \\
&\Rightarrow& g''(r) = \frac{1}{32} \big((1-r)^2 {}_2F_1(5/2,3/2;3;r)^2\\
&& -\,2 (1-r)^2\,{}_2F_1(5/2,3/2;2;r)\,{}_2F_1(3/2,5/2;4;r) \big)
\eea
We will show that ${}_2F_1(5/2,3/2;3;r)^2 \le \,{}_2F_1(5/2,3/2;2;r)\,{}_2F_1(5/2,3/2;4;r)$ for all positive $r$, implying that $g''(r)$ is negative in this region. We write out the two hypergeometric series explicitly:

\bea
{}_2F_1(5/2,3/2;3;r)^2 &=& \sum_{m,n=0}^\infty \frac{k_m k_n}{3^{\bar{m}} 3^{\bar{n}}} \; \mbox{, where~} k_n = \frac{(5/2)^{\bar{n}}(3/2)^{\bar{n}}}{n!} r^n\\
{}_2F_1(5/2,3/2;2;r)\,{}_2F_1(5/2,3/2;4;r) &=& \sum_{m,n=0}^\infty \frac{k_m k_n}{4^{\bar{m}} 2^{\bar{n}}}\\
&=&\sum_{m,n=0}^\infty \frac{k_m k_n}{3^{\bar{m}} 3^{\bar{n}}} \left(\frac{3}{3+m}\right) \left(\frac{2+n}{2}\right)
\eea
\bea
\label{bigSum} &=& \sum_{m=0}^\infty \frac{k_m^2}{3^{\bar{m}} 3^{\bar{m}}} \left(\frac{6+3m}{6+2m}\right) + \sum_{\substack{m,n=0\\m>n}}^\infty \frac{k_m k_n}{3^{\bar{m}} 3^{\bar{n}}} \left( \frac{3(2+n)}{2(3+m)} + \frac{3(2+m)}{2(3+n)} \right)\\
&\ge& \sum_{m=0}^\infty \frac{k_m^2}{3^{\bar{m}} 3^{\bar{m}}} + \sum_{\substack{m,n=0\\m>n}}^\infty \frac{2 k_m k_n}{3^{\bar{m}} 3^{\bar{n}}} = {}_2F_1(5/2,3/2;3;r)^2
\eea
where elementary methods can be used to show that the bracketed last term in eqn. (\ref{bigSum}) is at least 2 for any non-negative $m$ and $n$. This completes the proof of the lemma.
\end{proof}

\begin{figure}[htp]
\begin{center}
\includegraphics[clip,width=100mm,viewport=0 200 620 590]{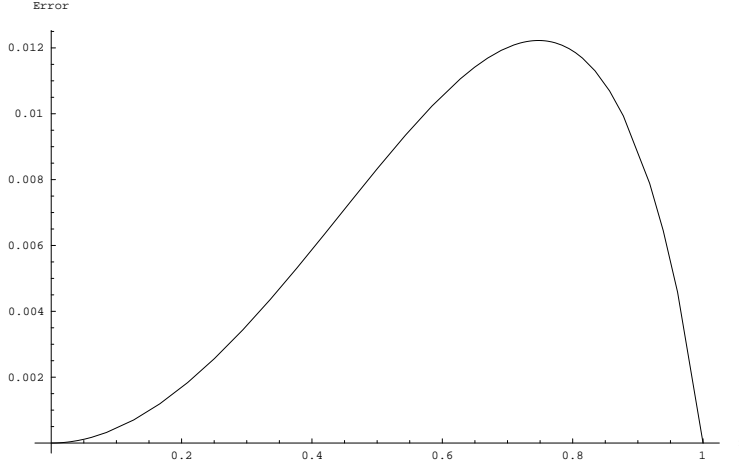}
\caption{Error in approximation to elliptic integral (\ref{integralIneq}) for $0 \le r \le 1$.}
\label{fig2}
\end{center}
\end{figure}


\section{Lipschitz constants}
\label{lipschitzAppendix}

This appendix contains derivations of the Lipschitz constants of the functions used for the concentration of measure results.

\begin{lipschitzLem*}
Let $S$ be an $n \times d$ matrix with $\|S\|_2=1$, and define $f(S) = \frac{1}{n}\|S\|_{tr}^2$. Then the Lipschitz constant $\eta$ of $f$ satisfies $\eta \le 2$.
\end{lipschitzLem*}

\begin{proof}
Let $k=\min(n,d)$. We have
\bea
\eta &=& \sup_{S,T} \frac{|f(S)-f(T)|}{\|S-T\|_2} = \sup_{S,T} \frac{|\,\|S\|_{tr}^2-\|S\|_{tr}^2\,|}{n \|S-T\|_2} \\
&=& \sup_{S,T} \left(\frac{\|S\|_{tr}+\|T\|_{tr}}{n}\right) \frac{|\,\|S\|_{tr}-\|S\|_{tr}\,|}{\|S-T\|_2} \\
&\le& \sup_{S,T} \left(\frac{\|S\|_{tr}+\|T\|_{tr}}{n}\right) \frac{\|S-T\|_{tr}}{\|S-T\|_2} \\
&\le& \sup_{S,T} \frac{\sqrt{k}\,(\|S\|_{tr}+\|T\|_{tr})}{n} \le 2k/n \le 2
\eea
The first inequality is a triangle inequality, and the second two are derived from

\be \label{normIneq} \|S\|_{tr} = \sum_{i=1}^k \sigma_i(S) \le \sqrt{k \sum_{i=1}^k \sigma_i^2(S)} \le \sqrt{k} \|S\|_2 \ee
which in turn uses a Cauchy-Schwarz inequality.
\end{proof}

\begin{lipschitzCubeLem*}
Let $S$ be a point on the $nd$-dimensional hypercube written down as an $n \times d$ $\{-1,1\}$-matrix, and let $f(S) = \frac{1}{n^2 d}\|S\|_{tr}^2$. Then the Lipschitz constant $\eta$ of $f$ (with respect to the Hamming distance) satisfies $\eta \le 4/nd$.
\end{lipschitzCubeLem*}

\begin{proof}
The proof is very similar to that of Lemma \ref{lipschitz}. As before, let $k=\min(n,d)$. We have

\bea
\eta &=& \sup_{S,T} \frac{|f(S)-f(T)|}{d(S,T)} = \sup_{S,T} \frac{1}{n^2d} \frac{|\,\|S\|_{tr}^2-\|S\|_{tr}^2\,|}{d(S,T)} \\
&\le& \sup_{S,T} \left(\frac{\|S\|_{tr}+\|T\|_{tr}}{n^2d}\right) \frac{\|S-T\|_{tr}}{\frac{1}{2}\|S-T\|_{1}} \\
&\le& \sup_{S,T} \frac{2 \sqrt{k}\,(\|S\|_{tr}+\|T\|_{tr})}{n^2d} \le 4k/n^2d \le 4/nd
\eea
where, extending inequality (\ref{normIneq}), we use $\|S\|_{tr} \le \sqrt{k} \|S\|_2 \le \sqrt{k} \|S\|_1$.
\end{proof}


\end{document}